\documentclass[10pt, conference, compsocconf]{IEEEtran}
\usepackage[normalem]{ulem}

\usepackage{amsmath}
\usepackage{epsfig,endnotes,ifpdf,pgfplots}
\usepackage[nocompress]{cite}
\usepackage[hyphens]{url}
\usepackage{moreverb}
\usepackage{multirow}
\usepackage{xcolor,colortbl}
\usepackage{array}
\usepackage{ragged2e}
\newcolumntype{P}[1]{>{\RaggedRight\hspace{0pt}}p{#1}}
\usepackage{mathptmx}
\usepackage{caption}
\usepackage{subcaption}
\usepackage{graphicx}
\graphicspath{{images/}}

\newcommand{\myfig}[5]
{
\begin{figure}[t]
\begin{center}
\ifpdf
\includegraphics[width=#4\linewidth]{#1}
\else
\includegraphics[width=#4\linewidth]{#1}
\fi
\end{center}
\vspace{#5}
\caption{#2}\label{#3}
\end{figure}
}

\newcommand{\myfigwide}[5]
{
\begin{figure*}[t]
\begin{center}
\ifpdf
\includegraphics[width=#4\linewidth]{#1}
\else
\includegraphics[width=#4\linewidth]{#1}
\fi
\end{center}
\vspace{#5}
\caption{#2}\label{#3}
\end{figure*}
}

\newcommand{\mysubfigtwo}[8]
{
\begin{figure}
        \centering
        \begin{subfigure}[b]{0.25\textwidth}
                \centering
                \includegraphics[width=\textwidth]{#1}
                \caption{#2}
                \label{#3}
        \end{subfigure}%
        ~ %add desired spacing between images, e. g. ~, \quad, \qquad etc.
        \begin{subfigure}[b]{0.25\textwidth}
                \centering
                \includegraphics[width=\textwidth]{#4}
                \caption{#5}
                \label{#6}
        \end{subfigure}
        ~ %add desired spacing between images, e. g. ~, \quad, \qquad etc.
        \caption{#7}\label{#8}
        \vspace{-15pt}        
\end{figure}
}

\newcommand{\mysubfigthree}[9]
{
\def\tempa{#1}
\def\tempb{#2}
\def\tempc{#3}
\def\tempd{#4}
\def\tempe{#5}
\def\tempf{#6}
\def\tempg{#7}
\def\temph{#8}
\def\tempi{#9}
\mysubfigthreecont
}

\newcommand{\mysubfigthreecont}[2]
{
\begin{figure*}
        \centering
        \begin{subfigure}[b]{0.3\textwidth}
                \centering
                \includegraphics[width=\textwidth]{\tempa}
                \caption{\tempb}
                \label{\tempc}
        \end{subfigure}%
        ~ %add desired spacing between images, e. g. ~, \quad, \qquad etc.
        \begin{subfigure}[b]{0.3\textwidth}
                \centering
                \includegraphics[width=\textwidth]{\tempd}
                \caption{\tempe}
                \label{\tempf}
        \end{subfigure}
        \begin{subfigure}[b]{0.3\textwidth}
                \centering
                \includegraphics[width=\textwidth]{\tempg}
                \caption{\temph}
                \label{\tempi}
        \end{subfigure}
        ~ %add desired spacing between images, e. g. ~, \quad, \qquad etc.
        \caption{#1}\label{#2}
        \vspace{-5pt}
\end{figure*}
}

\newcommand{\mysubfigthreebox}[9]
{
\def\tempa{#1}
\def\tempb{#2}
\def\tempc{#3}
\def\tempd{#4}
\def\tempe{#5}
\def\tempf{#6}
\def\tempg{#7}
\def\temph{#8}
\def\tempi{#9}
\mysubfigthreeboxcont
}

\newcommand{\mysubfigthreeboxcont}[2]
{
\begin{figure*}
        \centering
        \begin{tabular}{ccc}
        \begin{subfigure}[b]{0.32\textwidth}
                \centering
                {%
                \setlength{\fboxsep}{0pt}%
                \setlength{\fboxrule}{1pt}%
                \fbox{\includegraphics[width=\textwidth]{\tempa}}%
                }%
                \caption{\tempb}
                \label{\tempc}
        \end{subfigure}%
        &
        \begin{subfigure}[b]{0.32\textwidth}
                \centering
                {%
                \setlength{\fboxsep}{0pt}%
                \setlength{\fboxrule}{1pt}%
                \fbox{\includegraphics[width=\textwidth]{\tempd}}%
                }%
                \caption{\tempe}
                \label{\tempf}
        \end{subfigure}
        &
        \begin{subfigure}[b]{0.32\textwidth}
                \centering
                {%
                \setlength{\fboxsep}{0pt}%
                \setlength{\fboxrule}{1pt}%
                \fbox{\includegraphics[width=\textwidth]{\tempg}}%
                }%
                \caption{\temph}
                \label{\tempi}
        \end{subfigure}
        \end{tabular}
        \caption{#1}\label{#2}
        \vspace{-10pt}
\end{figure*}
}

\begin{document}

\hyphenation{Oca-sta}

\title{\Large \bf Ocasta: Clustering Configuration Settings For Error Recovery}
%\author{}
\author{
  {\rm Zhen Huang\hspace{0.4in} David Lie}\\
  {\it Department of Electrical and Computer Engineering}\\
  {\it University of Toronto}
% copy the following lines to add more authors
%  \and
% {\rm David Lie}\\
%  University of Toronto
} % end author
\date{} % no date on first page
\maketitle

\thispagestyle{empty}

\begin{abstract}

Effective machine-aided diagnosis and repair of configuration errors continues to elude computer systems designers.  Most of the literature targets errors that can be attributed to a single erroneous configuration setting.  However, a recent study found that a significant amount of configuration errors require fixing more than one setting together. To address this limitation, Ocasta statistically clusters dependent configuration settings based on the application's accesses to its configuration settings and utilizes the extracted clustering of configuration settings to fix configuration errors involving more than one configuration settings. Ocasta treats applications as black-boxes and only relies on the ability to observe application accesses to their configuration settings.

We collected traces of real application usage from 24 Linux and 5 Windows desktops computers and found that Ocasta is able to correctly identify clusters with 88.6\% accuracy.  To demonstrate the effectiveness of Ocasta, we evaluated it on 16 real-world configuration errors of 11 Linux and Windows applications. Ocasta is able to successfully repair all evaluated configuration errors in 11 minutes on average and only requires the user to examine an average of 3 screenshots of the output of the application to confirm that the error is repaired. A user study we conducted shows that Ocasta is easy to use by both expert and non-expert users and is more efficient than manual configuration error troubleshooting.

\end{abstract}


% \fi
\IEEEpeerreviewmaketitle

\section{Introduction}\label{sec:intro}

Configuration errors are a leading cause of failure and unavailability for desktop applications \cite{Ganapathi2004}.  Fixing such errors has essentially two steps: identifying the configuration settings causing the error, and replacing the faulty settings with values that fix the configuration error.  

To facilitate the first step, proposals in the literature have tried to pinpoint the time the configuration error first appeared~\cite{chronus}, used statistical anomaly detection to detect abnormal configuration settings~\cite{peerpressure,strider,glean}, or used white-box dynamic analysis to find the particular configuration setting that causes the application to execute an erroneous code path~\cite{confaid}.  Of these three approaches, only the last two try to identify the configuration setting that causes the error and even then, they only work if the error is the result of a single configuration setting.  Unfortunately, this can be a serious drawback since a recent study found that a significant number of configuration errors (14.9\%-34.7\%) require changing more than one configuration setting to fix~\cite{yin:sosp2011}, because some configuration settings are related. 

One example of related configuration settings is illustrated in Figure~\ref{fig:example1}: the number of ``Item'' settings should never exceed the value of {\tt Max Display} setting. Microsoft Word automatically maintains this relationship. For instance, if a user reduces the maximum number of recently accessed documents from the Preference menu, Microsoft Word not only reduces the value of {\tt Max Display} setting, but also deletes extra {\tt Item} settings. Consequently, if the user wants to undo the effect of reducing the maximum number of recently accessed documents, both the old value of {\tt Max Display} and the deleted {\tt Item} settings need to be recovered.

In this paper, we present a novel technique that uses hierarchical agglomerative clustering~\cite{DataMining} to identify clusters of related configuration settings, relying only on the ability to observe application accesses to its configuration store, and is thus language, binary and OS independent. We implemented this technique in Ocasta, which treats applications as black-boxes and is able to work on a wide range of applications and environments.  

%Identifying related configuration settings allows Ocasta to automatically fix a broader set of configuration errors than traditional systems. To the best of our knowledge, Ocasta is the first system that is capable of automatically fixing configuration errors involving both single and multiple configuration settings.

To evaluate the effectiveness of Ocasta, we collected traces of application usage from both Windows and Linux machines ranging from 18 to 76 days in length and then use Ocasta to identify clusters of related configuration settings in 11 different application in across 4 different OS flavors.  Using this data and 16 real-world configuration errors, we show that Ocasta's clustering is able to accurately identify 88.6\% of the clusters of related configuration settings.

To further evaluate Ocasta, we added a simple GUI-based configuration error repair tool that, with user input, uses the clustering information from Ocasta to automatically search for and fix settings causing configuration errors.  The Ocasta search tool requires the user to provide a GUI-action script that triggers the error, which it then uses to automatically search historical values of the clusters of configuration settings found by Ocasta for a fix.  A screenshot of the result is recorded after each search and the user is asked to select a screenshot that shows that the symptoms of the configuration have been treated.  

%We evaluate Ocasta against {\bf XX} real-world configuration errors and show that Ocasta's clustering allows Ocasta to find and fix all configuration settings necessary to repair the error, while only necessarily changing {\bf xxx} settings.

%\mysubfigsixbox{example1}{MS Word}{fig:example1}{example2}{Acrobat Reader}{fig:example2}{example3}{Evolution Mail}{fig:example3}{example4}{MS Paint}{fig:example4}{example5}{Internet Explorer}{fig:example5}{example6}{Explorer}{fig:example6}{Examples of dependency relationships among configuration settings}{}
\mysubfigthreebox{example1}{MS Word}{fig:example1}{example2}{Acrobat Reader}{fig:example2}{example3}{Evolution Mail}{fig:example3}{Examples of dependency relationships among configuration settings}{}

Configuration error repair in general is very hard and while Ocasta's proof of concept tool is able to fix the symptoms of all of our configuration errors, it cannot guarantee that the selected fix does not introduce new hidden errors, nor can it fix errors that do not have any visible symptoms.  In general, studies have shown that even trained humans may fail to fix configuration errors completely,  create new errors in the process troubleshooting or fixing an existing error, or have to resort to resetting the application back to its defaults to remove the symptoms of a configuration error~\cite{walfish:hotos2011}.  Our evaluation demonstrates that Ocasta's method for inferring related configuration settings broadens the range of errors automated configuration error repair tools can handle by providing with clustering information.  We believe that even when automated tools fail, the clustering information provided by Ocasta will still be valuable to human troubleshooters.

\vspace{2pt}
\noindent Our contributions are:
\begin{itemize}
\item We characterize the types and reasons of for relationships between configuration settings by manually inspecting and analyzing over 500 configuration settings across 11 applications.

\item We present the design and prototype implementation of Ocasta, which uses black-box statistical clustering of application behavior to identify related configuration settings.  Ocasta has been implemented on both Linux and Windows and evaluated on both systems using data collected from machines used by real people.

\item We further evaluate the usability of Ocasta's clustering with a proof-of-concept tool that given a set of actions that recreates a configuration error, automatically searches historical values of clusters of configuration settings for a fix.  We demonstrate the effectiveness of our tool against 16 real-world configuration errors.  We also provide a user study showing the effectiveness of Ocasta's configuration repair tool.

%We hav
%e collected traces of application usage from both Windows and Linux machines ranging from 18 to 76 days in length.  We present a study we conducted on these traces to ascertain both the severity of the collateral damage that can be caused by snapshot recovery.
%
%\item We present the design and prototype implementation of Ocasta, which provides unattended, efficient roll back recovery by that minimizes collateral damage.  Our prototype works on both Linux and Windows systems.  
%
%%We have deployed the Ocasta on personal computers and on machines in an undergraduate computing lab to gather traces of application behavior to evaluate Ocasta.
%
%\item  We combine our traces with real configuration errors to evaluate how well Ocasta does at recovery from such errors.  Ocasta is able to fix all configuration errors in under 7 minutes, requiring an average time of 2.5 minutes to find the error and requiring the user to examine fewer than 11 screenshots.  Collateral damage is reduced by an average of 72\%.  

\end{itemize}
\vspace{2pt}
We begin by studying relations between configuration settings and defining the problem solved by Ocasta in Section~\ref{sec:problem}.  We then describe Ocasta's high-level design in Section~\ref{sec:overview} and give implementation details in Section~\ref{sec:implementation}.  We describe how we collected our traces in Section~\ref{sec:data} and evaluate Ocasta in Section~\ref{sec:evaluation}.  Finally, we discuss related work in Section~\ref{sec:related} and conclude in Section~\ref{sec:conclusion}.

\begin{table}
\begin{tabular}{|p{1.3cm}|p{0.6cm}|p{1cm}|p{1cm}|p{1cm}|p{1.4cm}|}
\hline {\bf Name} & {\bf Days} & {\bf Reads } & {\bf Writes } & {\bf \# Keys} & {\bf TTKV Size} \\
%\multicolumn{1}{|c|}{\textbf{Name}} &
%\multicolumn{1}{|c|}{\textbf{Days}} &
%\multicolumn{1}{|c|}{\bf Reads} &
%\multicolumn{1}{|c|}{\bf Writes} &
%\multicolumn{1}{|c|}{\bf Deletes} &
%\multicolumn{1}{|c|}{\bf \# Keys} &
%\multicolumn{1}{|c|}{\bf TTKV Size (MB)} \\
\hline  Windows 7 & 42 & 6.76M & 67.72K & 4,611 & 85MB \\ % Windows 7 & 42 & 6,762,139 & 67,724 & 3,223 & 4,611 & 85 \\
\hline  Windows Vista & 53 & 3.46M & 20.5K & 14,673 & 29MB \\ % Windows Vista & 53 & 3,458,465 & 20,491 & 6,423 & 14,673 & 29 \\
\hline   Windows Vista-2 & 18 & 15.08M & 224.64K & 1,123 & 6.3MB \\ % Windows Vista-2 & 18 & 15,084,373 & 224,638 & 460 & 1,123 & 6.3 \\
\hline  Windows XP & 25 & 22.80M & 311.9K & 14,667 & 24MB \\ % Windows XP & 25 & 22,797,779 & 311,897 & 1,516 & 14,667 & 24 \\
\hline  Windows XP-2 & 32 & 26.76M & 268.96K & 19,501 & 46MB \\ % Windows XP-2 & 32 & 26,763,343 & 268,964 & 2,894 & 19,501 & 46 \\
\hline  Linux-1 & 25 & 91.52K & 3.34K & 1,660 & 6MB \\ % Linux-1 & 25 & 91,521 & 3,335 & 0 & 1,660 & 6 \\
\hline  Linux-2 & 84 & 8.15K & 0.48K & 35 & 0.1MB \\ % Linux-2 & 84 & 8,147 & 477 & 0 & 35 & 0.1 \\
\hline  Linux-3 & 46 & 52.41K & 0.44K & 706 & 0.7MB \\ % Linux-3 & 46 & 52,406 & 435 & 0 & 706 & 0.7 \\
\hline  Linux-4 & 64 & 507.07K & 5.43K & 751 & 6.4MB \\ % Linux-4 & 64 & 507,069 & 5,429 & 0 & 751 & 6.4 \\
\hline
\end{tabular}
\caption{Summary of trace statistics.  The traces on the Linux machines are aggregated by users instead of machines.  We only list statistics for users whose data we use in the evaluation of this paper.  The last column gives the size of the TTKV at the end of the trace.  For Linux-2, Linux-3 and Linux-4, the TTKV only stores keys from the application-file logger.}
\label{tbl:datacollection}
\vspace{-10pt}
\end{table}

\section{Problem Definition}
\label{sec:problem}

Similar to relationships between program variables~\cite{Lu2007}, relationships between configuration settings are a common, though not often documented phenomenon that applications exhibit. We begin by describing 3 representative examples of related configuration settings that we found by manually inspecting over 500 configuration settings that were accessed by 11 different Windows and Linux applications in our traces (trace statistics given in Table~\ref{tbl:datacollection}).  %We also broadly categorize the related configuration settings into categories: those with a hierarchical dependency, where one setting determines the validity, arrangement, or meaning of a group of other settings, and correlated settings, where the settings are mutually related.  We note that strictly speaking, settings in these categories do not always have to be changed together.  However, in most cases changes in one setting will require changes in one or more other related settings to allow the application to function properly.

\vspace{6pt}
%\noindent {\bf Dependent Configuration Settings:}
In Figure~\ref{fig:example1}, to control the number of documents listed in the recently opened documents list in Microsoft Word, {\tt Max Display} limits the number of document names stored in the {\tt Item} settings (e.g. {\tt Item 1}, {\tt Item 2}).  In Figure~\ref{fig:example2}, Acrobat Reader uses {\tt InlineAutoCompelete} to determine whether to enable the ``auto complete'' feature when user fills a form, while {\tt RecordNewEntries} and {\tt ShowDropDown} specify how the ``auto complete'' feature works, including whether to record user-entered data and whether to display the list of previously recorded data in a dropdown box. Finally, in Figure~\ref{fig:example3}, Evolution will automatically mark an opened email as ``seen'' after an email has been opened by the user for the time interval specified by the value of {\tt mark\_seen\_timeout}, but only when {\tt mark\_seen} is set to ``true''.  These examples illustrate that related configuration settings exist when one or more settings controls the validity or meaning of another group of settings.  

\iffalse
\noindent {\bf Correlated Settings:}
The settings in Figure \ref{fig:example5} describe different properties such as name, style, and address of the 1st URL link displayed in the favorite links bar in Internet Explorer, with {\tt DisplayName}, {\tt DisplayMask}, {\tt FeedUrl}, respectively. In Figure \ref{fig:example6}, {\tt Maximize} specifies whether the Windows Explorer Shell should display an image file in Maximized or Windowed mode, and {\tt Bounds} specifies the bounds in which the image file should be displayed.  Unlike hierarchically dependent settings, these settings are equally or mutually dependent.
\vspace{6pt}
\fi

Because related configuration are designed to work together, applications are likely to update related configuration settings together, in order to satisfy their relation as illustrated in our 3 examples. In addition, users tend to change related configuration settings together. For example, a user will probably set the value of {\tt mark\_seen\_timeout} and change the value of {\tt mark\_seen} to ``true'' together, in order to enable Evolution to automatically mark an opened email. In contrast, independent configuration settings are unlikely to be changed together. Based on this intuition, Ocasta identifies the relations among configuration settings by observing the access correlations among them and uses hierarchical agglomerative clustering to group together configuration settings based on access correlations.

\paragraph{Limitations} Ocasta has several limitations. First, independent configuration settings can be accidentally updated simultaneously and cause the hierarchical agglomerative clustering algorithm that Ocasta uses to incorrectly identify them as dependent. Similarly, partial update of dependent settings may be legal in some cases causing Ocasta to incorrectly infer that related settings should be in separate clusters. Ocasta's clustering can be tuned to handle such cases, but this tuning may require some manual intervention.  Ultimately, Ocasta can only perform as well as the quality and amount of data available to it.  Second, Ocasta must be able to intercept and record accesses to the individual keys where the application stores its persistent settings.  We have implemented and tested such capabilities for OS-provided key-value stores like the Windows Registry and GConf in Linux.  While many applications use OS-provided stores, some applications use their own files to store configurations.  Thus we have also implemented custom parsers for several common file formats, such as XML, JSON, PostScript, INI and plain text.% and give details on these in Section~\ref{sec:implementation}.

Ocasta's proof-of-concept error repair tool has some additional limitations.  First, a fix for the configuration error must exist in the application's recorded history.  Our tool cannot fix applications that have always been misconfigured or where the configuration error arose due to a change in an external dependency.  Second, the configuration error must occur deterministically, because our tool only performs one trial execution per historical cluster value in its search.  Finally, because the user must be able to identify a fixed application from its screenshot, the configuration error must be visually observable on the display.  

\section{Overview}\label{sec:overview}

\subsection{Clustering configuration settings}
\label{sec:clustering}

%Ocasta reduces collateral damage by rolling back only the keys that are needed to fix a configuration error.  However, rolling back too few keys may corrupt the application's configuration state by breaking dependencies among related keys.  Thus, it is critical that Ocasta identify clusters of related keys.

Ocasta improves configuration troubleshooting and repair by heuristically identifying clusters of related configuration settings.  Ocasta abstracts configurations into key-value pairs, with the key being the name of the configuration setting and the value being the content of the setting.  As we see in Section~\ref{sec:implementation}, many application configurations naturally fit into this abstraction.  

It is important that the clusters of configuration settings that Ocasta extracts from observing application behavior be accurate.  On one hand, extracting {\em undersized} clusters can create clusters that do not contain all the configuration keys necessary to fix a configuration error.  Even worse, attempting to fix an error with an undersized cluster can, in some cases, break dependencies between configuration settings, leading to a non-working application configuration.

On the other hand, extracting {\em oversized} clusters causes unrelated configuration settings to be clustered together, and can lead to extraneous configuration changes when trying to repair errors.  As an extreme example, repairs that reset an application configuration back to its defaults, or copy a configuration from a previous snapshot or a different user, essentially treat the application's configuration as a single, large, oversized cluster.

Ocasta uses the property that related configuration keys are much more likely to be modified together than unrelated keys to infer which keys are related.  To determine whether keys have been modified together, Ocasta uses a sliding time window and considers all keys written within the window to have been modified together.  Ocasta uses a default sliding window of 1 second, which can be increased if needed by the user.  Some keys are modified very frequently, so the chances of such a key being modified concurrently with unrelated keys is high.  Consequently, Ocasta only clusters together keys that are often modified together, but rarely modified individually on their own or with other keys.  To do this, we define a {\em correlation} metric between each pair of keys:
$$
  Correlation = \frac{|A \cap B|}{|A|} + \frac{|A \cap B|}{|B|}
$$
$A$ and $B$ denote the set of all writes to keys {\tt A} and {\tt B} respectively, and the intersection of $A$ and $B$ denotes the set of writes where both keys were written together.  The correlation metric is maximized at 2 when both keys are always modified together and minimized at zero when both keys are never modified together.  The larger the correlation, the more related the pair of keys. Note that the correlation is only defined when both keys have a non-zero number writes.  Since Ocasta assumes that the application worked initially, any key that has not been modified from its initial value cannot cause a configuration error, and is thus excluded from Ocasta's search for a configuration fix.

Hierarchical agglomerative clustering~\cite{DataMining} takes as input a set of points, distances between each pair of points, and a linkage criterion that defines how distances between clusters are computed.  It then iteratively merges clusters together, forming a hierarchy with larger clusters at the top of the hierarchy.  In Ocasta, we use the ``maximum linkage criterion'', which defines the distance between a pair clusters as the maximum distance between any two keys across the clusters. Hierarchical clustering has the advantage over other types of clustering, such as k-means or centroid-based clustering, in that it does not require  the number of clusters to be specified in advance. To perform hierarchical clustering, distances need to be smaller as keys become more related, so we use the inverse of our correlation metric as the distance for Ocasta's clustering. To decide when to stop clustering, Ocasta provides a tune-able {\em threshold}, which defines the maximum distance between any two clusters. By default, Ocasta uses a threshold equivalent to a correlation value of 2 (i.e. a distance of 0.5), which only clusters keys that are always modified together.  If the user finds that configuration repair fails due to undersized clusters, she may decrease the threshold to allow Ocasta to cluster together keys that are modified together most of the time.

Like any black-box heuristic, Ocasta can fail under certain circumstances, particularly for configuration settings that have had very few modifications from which Ocasta can learn.  For example, the user may modify several unrelated settings at once, causing the application to store those changes together into its configuration store.  Unless, these settings are later modified separately, Ocasta will incorrectly infer that they are related, resulting in an oversized cluster.  Similarly, it is possible that a user makes a single change to an application that causes a change to only one level of hierarchically dependent configuration keys.  For example, she may disable the feature completely, which would only change the higher-level key, modify the lower-level keys without changing the higher-level key, or only modify a subset of the lower-level keys.  Again, if this was the only instance of modifications to the key, then Ocasta may infer an undersized cluster that separates related keys from each other into different clusters.  While only using black-box information makes Ocasta more broadly applicable, Ocasta can only work with the information it observes and as a result, can be misled when there is inadequate history for its clustering to work.

\subsection{Automated repair}
%When invoked in recovery mode, Ocasta requires the user to provide a trial that demonstrates the configuration error.  

Ocasta's automated repair tool uses the clustering information to aid the user in fixing configuration errors.  For example, configuration error \#15, described in Table~\ref{tbl:misconfig}, causes the menu bar to disappear when certain PDF documents are opened in Acrobat Reader.  To use Ocasta, the user must first create a {\em trial}, which tells Ocasta how to recreate the error and makes the symptoms of the error visible on the screen.  For example, in the case of error \#15, the user starts Acrobat Reader and uses it to open the PDF document that causes the error.  Since the menu bar disappears once the document is opened, the error is visible on the screen.  The user thus ends the trial with the menu bar missing and document open on the screen.  Ocasta records the UI actions the user made in the trial and automatically extracts the identity of the application or applications that were used.  
%
%After Ocasta has recorded the trial, it 

Ocasta's repair tool then asks the user to specify an optional {\em start time} and an optional {\em end time}.  The start time is the earliest time the user believes the configuration error could have been introduced, and allows Ocasta to limit how far back in time it searches for the cluster that causes the error, which we call the {\em offending cluster}.  If the user doesn't specify a bound, Ocasta will search all the cluster versions in the recorded history of the application.  The end time is the latest time the user believes the configuration error could be introduced and should roughly coincide with time the configuration error is first discovered.  This is useful if the user might have tried to fix the error themselves and thus may have made spurious configuration changes that might slow down the search.  If the user does not specify an end time, Ocasta uses all recorded values up to the end of the recorded history.

In some cases Ocasta can identify a large number of clusters in an application (as many as 220 in our measurements).  As a result, recovery will be significantly faster if Ocasta sorts clusters so that the ones that are likely to be configuration clusters are checked before the ones that are likely to be non-configuration clusters.  We use the intuition that changes to configuration settings should be infrequent because for them to change, the user must explicitly modify a configuration setting, which also happens infrequently. Ocasta thus sorts the clusters by the number of times they have been modified over the application's history. 

Ocasta then executes the user-provided trial on the historical values of the clusters by rolling back an entire cluster of configuration settings at a time and running the trial in a sandbox, which prevents the execution to leave any persistent changes.  
Ocasta can be configured to perform either a breadth-first (BFS) or depth-first (DFS) search on the historical values of each cluster.  In DFS, Ocasta executes the trial on all the historical values of a cluster before moving onto the next cluster.  In BFS, Ocasta executes the latest historical value of each cluster before moving onto the next historical value.  DFS works well if Ocasta's sort algorithm successfully prioritizes the offending cluster early in the sort, while the BFS algorithm provides performance that is less influenced by how well the sort worked.  

After each trial execution, the tool takes a screenshot.  Ocasta discards the screenshot if it is identical to either the erroneous screenshot or any previous  screenshots it has recorded.  The user can periodically check on the recorded screenshots recorded to see if any of them display a fixed configuration.  When she see a fixed configuration, Ocasta permanently rolls back the cluster to its corresponding value and returns back to recording mode.  A video demonstrating the use of Ocasta is available online for viewing~\footnote{\url{http://youtu.be/aRvJlTj-0F0}}.
\section{Implementation}
\label{sec:implementation}

In this section we describe implementation details of Ocasta's prototype. Ocasta works on both Windows and Linux.  Ocasta supports applications that use the Windows registry or the GConf configuration system, as well as applications that store configuration state in XML, JSON, PostScript, INI and plain text files.  We describe the implementation of the Ocasta time travel key-value store, the logger, as well as the clustering and repair components of Ocasta.

\subsection{Time travel key-value store}

% Reads removed
Ocasta records configuration key-value activity in a time travel key-value store (TTKV).  We implemented Ocasta's TTKV using Redis, a commonly used key-value store~\cite{redis}.  Redis maps each key in the application to a record that contains the number of writes and deletions, as well as a list of historical values of the key including timestamps.  A special type of value is used to represent deletions of the key, which are also recorded in the value history.

During regular application use, Ocasta's loggers (described in the next section) intercept accesses by applications to their configuration store and record information about these accesses in the TTKV.  Ocasta then uses the information stored in the TTKV to compute the clustering information for the keys.  In addition, Ocasta's configuration error repair tool uses historical values in the TTKV when performing its search for a configuration error fix.

\subsection{Logger}
The primary purpose of the logger is to intercept accesses an application makes to its persistent storage and abstract those into key-values that can be stored into the TTKV.  As a result, the logger is necessarily dependent on the way the application stores its application state.  Below we detail the implementation of Ocasta loggers for the Windows registry, GConf configuration system, and various file formats used by the applications we tested.

%Since Windows registry, Gconf configuration system, and configuration files are all accessed and writed in different manners, and there is no uniform way to intercept access and write to all of them, we implemented different versions of logger specifically for each of them.

%Logger intercepts key-value store API calls and file system API calls invoked by applications and stores the relevant information on each API call

\subsubsection{Windows registry}
The Windows registry is a key-value store provided by the Windows OS. Applications write keys in the Windows registry using a well-documented API provided by the OS.  We implemented the Windows registry logger as a user-space shared library.  To intercept registry API calls made by applications, we use the Windows debug APIs to inject the shared library into Explorer, the Windows shell. Once injected into Explorer, the shared library intercepts each Windows registry API by hooking the first five bytes of the instructions of the API call in a way similar to Detours~\cite{detour}.   The shared library also injects itself into new processes created by the process it is loaded into by intercepting the Windows API call that creates new processes.  Virtually all regular applications are started via the Explorer shell, which implements all the common methods for starting applications such as the Start Menu, desktop shortcuts, taskbar shortcuts, or double-clicking an executable in a folder.  As a result, the Ocasta logger is able to monitor every application a user uses.  We note that the Windows registry logger only captures registry activity by user applications, not by system services or the Windows kernel, so our current prototype cannot fix configuration errors in those components.

\subsubsection{GConf configuration system}
The GConf configuration system, commonly found on Linux systems, implements the handlers for its APIs in a shared library. We used the standard approach of intercepting shared library calls on Linux by using the {\tt LD\_PRELOAD} environment variable to load our own shared library into the address space of every process.  Our library exports a set of shared library calls that is identical to the set of shared library APIs exported by the GConf shared library.  By specifying our library in the {\tt LD\_PRELOAD} environment variable, our library is always loaded before the GConf library and thus all calls to those APIs will invoke our functions, which will then subsequently call the real functions in the GConf shared library after logging the events to the TTKV.

%has the priority of being loaded before the real shared library that implemented the APIs to be intercepted, the application to which the shared library is injected will call the APIs exported by our injected shared library instead of GConf's shared library.

%Similar to the shared library we used on Windows to intercept Windows registry API calls, we developed a shared library on Linux to intercept Gconf API calls made by applications. We modify the system environment variable LD\_PRELOAD to inject the shared library upon the creation of a process.
%we developed a kernel module on Linux to intercept file system calls to record access and write made by these applications to the file system. From our analysis on the data collected from the 24 Linux desktop machines, we found out that all the applications we examined follow a common practice of storing their key-value pairs in text files under some hidden directories (i.e. directory name starting with a dot) of a user's home directory. Almost all these text files are of some standard text formats. For example, OpenOffice uses XML file format for its configuration files and Chrome browser uses JSON file format for its configuration files. We also found that the text formats utilized by applications that do not use standard text formats are usually very simple. For example, Firefox uses a "key=value" like format to store many of its key-value pairs, while it uses some slightly modified format of INI file format to store other key-value pairs.

\subsubsection{Application-specific file formats}

\iffalse
\begin{table}[tb]
\begin{center}
%\resizebox{1\columnwidth}{!}{
%\scalebox{1}{
%\begin{tabular}{|l|r|l|l|}
\begin{tabular}{|l|p{0.7\linewidth}|}
\hline
\multicolumn{1}{|c|}{\bf File Type} & 
\multicolumn{1}{|c|}{\bf Application} \\
\hline  {JSON} & Chrome browser \\
\hline  {XML} & OpenOffice, Notepad++, VirtualBox, TrueCrypt \\
\hline  {INI} & Firefox browser, Audacity, TexMaker \\
\hline  {Plain Text} & VLC \\
\hline PostScript & Acrobat Reader \\
\hline
\end{tabular}
%}
\caption{Application-specific file formats for several common applications.}
\label{tbl:configfile}
\end{center}
\vspace{-20pt}
\end{table}
\fi

Applications that don't use OS-provided key-value storage facilities such as the Windows Registry or GConf generally implement their own file-based key-value store.  We conducted a small study on the common file formats used for configuration storage and found applications generally use standard file format: JSON, XML, PostScript, or one of two key-value lists that both had the format ``$key=value$'', which we called INI if it is hierarchical and plain text if it is flat. 

We elide the details of the implementation of our application-specific file parsers for the sake of space.  One inherent shortcoming of Ocasta when dealing with application-specific file formats is that applications typically read the entire file into an in-memory key-value store.  The applications then perform writes on the in-memory store and flush the in-memory store back to disk.  To infer which keys are changed, Ocasta compares the files before and after each flush.  In practice, we observe that applications typically flush their in-memory store after each key modification to guarantee persistence, but if they do not, Ocasta will not be able to tell if a key was modified several times between flushes.  As shown in Section~\ref{sec:evaluation}, despite the coarser level of information available to Ocasta for applications that use application-specific files, Ocasta is still able to offer good clustering performance for these applications.

\subsection{Ocasta clustering and repair tool}
\label{sec:recovery_manager}

Ocasta's clustering algorithm is based on an open source clustering library\cite{ClusteringSoftware}. However, the hierarchical clustering API provided by this library does not allow a cluster threshold to be used to restrict clustering. Hence, we added functionality to prune the results returned by the hierarchical clustering API according to a specified threshold. 

Ocasta's repair tool has three  main components -- a UI record and replay tool, which records the user-provided trial and re-executes it on the application, a screenshot tool, which takes and records screenshots of the application and a controller, which coordinates the entire recovery search.  We have implemented the repair tool on both Windows and Linux.  To save time and effort, we made judicious use of various open-source libraries and packages for recording UI actions, as well as capturing and manipulating screenshots.

A limitation with our current implementation of the repair tool is that it deterministically replays trials and thus does not guarantee the same trial can be replayed correctly across different configuration settings. A robust adaptive replay can probably address this limitation, but the current focus of our work is to demonstrate the benefits of clustering. Nonetheless, we found our repair tool works well in our evaluation and user study.

\section{Data collection}
\label{sec:data}

We deployed Ocasta on 24 Linux desktop computers running Debian 6 and 5 Windows desktop computers. Ocasta intercepts and records reads, writes and deletions of settings into application configuration stores such as the Windows registry, GConf database and application configuration files.  Configuration settings are abstracted into keys and stored into a key-value store called the Time Travel Key Value Store (TTKV).  Table \ref{tbl:datacollection} summarizes the characteristics of the traces from these deployments, which we use in this paper.  The period of deployments range from one month to over two months. All the computers were actively used during the deployment.

%\subsection{Linux desktop computers}
All the Linux desktop computers are from four undergraduate computing laboratories administrated by our department. To reduce bias in the selection of the computers, we choose 6 computers from each laboratory. These computers are used mainly on site by undergraduate students for their course work, and remotely by graduate students and faculty members in our department.  This study was approved by our institutional ethics review board.

%Data from these users was anonymized by hashing and uploaded to our servers nightly. 

%All 24 machines have essentially the same hardware configuration -- Intel Core2 Quad CPU, 8GB memory, and 12GB hard drive for the root partition. Home directories are stored on a shared NFS server and all machines have the same user accounts configured.

Because these machines are shared among many users, we link usage of applications by the same user regardless of what machine they are using -- traces from one machine by a particular user will be combined with traces from another machine by the same user.  Our ethics review board required us to only instrument a fraction of the computers in any one lab to give students who did not wish to participate in the study ample opportunity to select an uninstrumented machine.  Unfortunately, this meant that we only got a sampling of user-behavior since a student would not be likely to use an instrumented machine every time they were in the lab.

%\subsection{Windows desktop computers}
The 5 Windows desktop computers are personal computers used by four graduate students and one faculty member. They run a variety of Windows OS including Windows 7, Windows Vista, and Windows XP.

%Most Windows computers use Intel Core2 Dual CPU, and one Windows machine use Intel Core2 Quad CPU. One Windows machine is equipped with 4GB memory, and each of the other Windows computers are equipped with 2GB memory. The size of the hard drives of these computers range from 100GB to 200GB.

%\subsection{Limitations}
%Although we collect usage data from a variety of computers, we warn the readers that all the users of these computers are from an academic setting. Therefore we do not recommend the reader to generalize our results, albeit we believe our technique would perform well in different settings.

%\myfigwide{Ranking}{Speedup of Ocasta versus backwards search in time.  The height of each bar is computed by dividing the number of rollbacks the backwards search requires by the number of rollbacks Ocasta requires.  }{fig:rankingcomparison}{0.7}{-12pt}

\begin{table}
\begin{tabular}{|P{1.9cm}|p{1.8cm}|p{0.6cm}|p{1cm}|p{1.25cm}|}
\hline {\bf Application} & {\bf Description} & {\bf \#Keys} & {\bf \#Clusters} & {\bf \%Accuracy} \\ % \#Oversized
\hline MS Outlook & E-mail Client & 182 & 33/82 & 97.0\% \\%98.8\% \\ % 1   
\hline Evolution Mail & E-mail Client & 183 & 18/65 & 38.9\% \\%83.1\% \\ % 11
\hline Internet Explorer & Web Browser & 33 & 9/12 & 66.7\% \\%75.0\% \\ % 3
\hline Chrome Browser & Web Browser & 35 & 1/34 & 100\% \\ % 0
\hline MS Word & Word Processor & 143 & 18/110 & 100\% \\ % 0
\hline GNOME Edit & Word Processor & 10 & 1/7 & 0.0\% \\%85.7\% \\ % 1
\hline MS Paint & Image Editor & 66 & 2/8 & 50.0\% \\%87.5\% \\ % 1
\hline Eye of GNOME & Image Viewer & 5 & 0/5 & N/A \\ % 0
\hline Acrobat Reader & Document Reader & 751 & 120/550 & 95.8\% \\%99.1\% \\ % 5
\hline Explorer & Windows Shell & 298 & 32/91 & 84.4\% \\%94.5\% \\ % 5
\hline Windows Media Player & Media Player & 165 & 21/41 & 90.5\% \\%95.1\% \\ % 2
\hline {\bf Total} & N/A & 1,871 & 255/1,005 & 88.6\% \\
\hline
\end{tabular}
\caption{Applications and their clusters Identified by Ocasta. In column \#Clusters, we show two numbers: the number of clusters that have more than one configuration setting, followed by the number of all clusters.}
\label{tbl:clustering}
\end{table}

\begin{table*}[tbh]
\begin{center}
\begin{tabular}{|c|l|p{1in}|l|p{0.90\columnwidth}|}
\hline
\multicolumn{1}{|c|}{\textbf{Case}} &
\multicolumn{1}{|c|}{\textbf{Trace}} &
\multicolumn{1}{|c|}{\textbf{Application}} &
\multicolumn{1}{|c|}{\textbf{Logger}} &
\multicolumn{1}{|c|}{\textbf{Description}}  \\
\hline  1 &  Windows 7 & MS Outlook & Registry& User is unable to use Navigation Panel.  \\
\hline  2 &  Windows 7 &  MS Word & Registry& User loses the list of recently accessed documents.   \\
\hline  3 &  Windows 7 & Internet Explorer & Registry & Dialog to disable add-ons always pops up. \\
\hline  4 & Windows Vista & Explorer & Registry & ``Open with'' menu does not show installed applications that can open .flv file. \\
\hline  5 & Windows XP & Windows Media Player & Registry& Caption is not shown while playing video.  \\
\hline  6 & Windows XP & MS Paint & Registry& Text tool bar does not pop up automatically when entering text.   \\
\hline  7 & Windows XP & Explorer & Registry& Image files are always opened in a maximized window.   \\
\hline  8 &  Linux-1 &  Evolution Mail &  GConf & Evolution Mail starts in offline mode unexpectedly.  \\
\hline  9 &  Linux-1 & Evolution Mail & GConf & Evolution Mail does not mark read mail automatically.   \\
\hline  10 & Linux-1 & Evolution Mail &  GConf &  Evolution Mail does not start a reply at the top of an e-mail.  \\
\hline 11 & Linux-1 & Image Viewer & GConf & User is unable to print image files. \\
\hline 12 & Linux-1 & Text Editor & GConf & User is unable to save any document. \\
\hline 13 & Linux-2 & Chrome Browser & File & Bookmark bar is missing. \\
\hline 14 & Linux-2 & Chrome Browser & File & Home button is missing from the tool bar. \\
\hline 15 & Linux-3 & Acrobat Reader & File & Menu bar disappears for certain PDF document. \\
\hline 16 & Linux-4 & Acrobat Reader & File & Find box is missing from the tool bar. \\

%\hline  3 &  Windows XP & OUTLOOK & Registry& Duplicate personal folders    \\

\hline
\end{tabular}
\caption{Real configuration errors used in our evaluation. \label{tbl:misconfig}}
\end{center}
\vspace{-10pt}
\end{table*}

\section{Evaluation}\label{sec:evaluation}

We evaluate 3 aspects of our Ocasta prototype.  First, we evaluate the accuracy of the clusters that Ocasta extracts. Second, we evaluate the effectiveness and performance of Ocasta, and the benefits of using clustering at recovering from configuration errors. Finaly, we perform a user study to evaluate how easy it is for a user to generate a trial, identify the screenshot showing a fixed application, and use Ocasta in general.  All Windows experiments were performed on an Intel Core Duo Dual-Core laptop with 2 GB of memory running Windows 7 and all Linux experiments were performed on a Intel Core 2 Quad-Core desktop with 4 GB of memory running Debian 6. We used 11 popular desktop applications in our evaluations, as listed in Table~\ref{tbl:clustering}.

\subsection{Clustering Analysis}
%Similar to the approach taken by previous work \cite{2007lu, 2004li}, we have manually examined 100 randomly selected clusters.
To evaluate the accuracy of Ocasta's clustering algorithm, we manually examined all 255 clusters, each of which contains more than one configuration setting, across all applications used in our evaluations. First, we try to confirm whether configuration settings are correlated by examining their names and values. We identify relations of configuration settings from their hierarchical names \cite{glean} and verify their relations from their values. Second, we individually change configuration settings in a cluster and check whether the corresponding application runs properly after the change. We conservatively consider a cluster as correctly identified if and only if there is a dependency relationship among every configuration setting of the cluster. 

As a result, we define an {\it oversized cluster} as a cluster that contains one or more extra configuration settings that are not related with the other configuration settings in the cluster, and an {\it undersized cluster} as a cluster that does not contain one or more configuration settings that are related with the configuration settings in the cluster.

%For each cluster, we identified whether there is a dependency relationship among any subset of the configuration settings belong to the cluster. 

%We show the accuracy of Ocasta's clustering algorithm in Table~\ref{tbl:clustering}. For each application, we compute the ratio of oversized clusters over the number of clusters with more than one setting.  We also compute an total ratio by dividing the total number of oversized clusters over the total number of clusters with more than one setting across all applications.  

%The result illustrates that the ratio of oversized clusters is overall low, at 11.3\% on average, indicating that Ocasta correctly identifies clusters in 88.7\% of the cases. Except for four applications (Evolution Mail, Internet Explorer, Text Editor, and MS Paint) that have a very small number of clusters (smaller than 20) and a small number of configuration settings, all applications have a ratio of oversized clusters smaller than 15\%. We elaborate on our findings below.

We show the accuracy of Ocasta's clustering algorithm in Table~\ref{tbl:clustering}. For each application, we compute the ratio of correctly identified clusters with more than one setting over the total number of clusters with more than one setting. The result illustrates that Ocasta has a high accuracy of identifying clusters with more than one setting, 72.3\% on average (mean accuracy among all applications) and 88.6\% overall (ratio of the total number of correctly identified clusters to the total number of clusters across all applications). Except for four applications (Evolution Mail, Internet Explorer, Text Editor, and MS Paint) that have a very small number of clusters (smaller than 20) and a small number of configuration settings, Ocasta accurately identified clusters with more than one setting in 94\% of the cases. We elaborate on our findings below. 

%We note that using this measure, applications with a larger number of clusters will have a larger effect on the overall ratio.

%Like other statistics, the larger the sample size the more accurate the result. Hence the ratio of oversized clusters of each application should not be treated equally, since each application has a different number of clusters of configuration settings. As a result, we use the weighted average ratio of oversized clusters, using the ratio of the number of clusters of each application to the total number of clusters of all applications as the weight of the ratio of oversized clusters of each application. 
%

\vspace{3pt}
\paragraph{Oversized Clusters} The majority of the incorrectly identified clusters are oversized clusters, which are caused by two major sources. First, Ocasta is limited to using a minimum of one second as the sliding time window.  This is because the trace collection infrastructure only records the update time of configuration settings to the precision of the nearest second. Although the 1-second sliding time window works well for most applications, one second is long enough for an application to update more than one group of dependent configuration settings. For example, one oversized cluster of Evolution Mail contains six groups of dependent configuration settings. Second, some configuration settings may be updated simultaneously as the result of software updates, in which case even independent configuration settings could be updated together.

Oversized clusters can cause unnecessary configuration settings to be changed when attempting to fix configuration errors. As a result, we want to minimize the number of oversized clusters and the number of extra configuration settings in oversized clusters. To achieve that, we examined all 17 oversized clusters of the four applications with the highest ratio of oversized clusters. We found that 11 of the oversized clusters are composed of several groups of dependent configuration settings and that the remaining 6 of them have one extra configuration setting in them. This indicates that most of the oversized clusters are probably caused by using a 1-second sliding time window and could potentially have been eliminated if our trace collection infrastructure had recorded key modification times at a finer granularity. 

\paragraph{Undersized Clusters} Ocasta's clustering algorithm can also cause undersized clusters if dependent configuration settings are not always updated together. Undersized clusters can cause failures in fixing configuration errors, since dependent configuration settings are not changed together, or leave configuration settings in an inconsistent state that can cause application misbehavior.  In the next section, we describe how out of 16 injected errors, Ocasta is able to fix all but 2 using the default clustering threshold of 2 and window size of 1 second.  The 2 unfixed errors are a result of undersized clusters, which we were able to correct by tuning of the clustering threshold and window size.  We did not observe any application crashes or misbehavior during the hundreds of clusters that were changed during the trials executed by Ocasta to fix these errors.  
\vspace{6pt}

\begin{table}[tb]
\begin{tabular}{|p{0.5cm}|p{0.7cm}|p{0.6cm}|p{1.5cm}|p{0.9cm}|p{0.7cm}|p{0.8cm}|}	
\hline {\textbf{Case}} & {\textbf{Cl.Size}} & {\textbf{Trials}} & {\textbf{Time(mm:ss)}} & {\textbf{Screens}} & {\textbf{Ocasta}} & {\textbf{NoClust}} \\
%\multicolumn{1}{|c|}{\textbf{Case}} &
%\multicolumn{1}{|c|}{\textbf{Cl.Size}} &
%\multicolumn{1}{|c|}{\textbf{Trials}} &
%\multicolumn{1}{|c|}{\textbf{Time(mm:ss)}} & 
%\multicolumn{1}{|c|}{\textbf{Screens}} &
%\multicolumn{1}{|c|}{\textbf{Ocasta}} &
%\multicolumn{1}{|c|}{\textbf{Ocasta}} \\
\hline  1 & 2 & 15 & 0:30/6:00 & 5 & Y & Y\\
\hline  2 & 8 & 2 & 0:34/1:01 & 1 & Y & N\\
\hline  3 & 2 & 14 & 4:16/5:24 & 11 & Y & Y\\
\hline  4 & 3 & 33 & 3:02/8:57 & 1 & Y & N\\
\hline  5 & 4 & 60 & 5:36/28:40 & 1 & Y & Y\\
\hline  6 & 8 & 8 & 3:04/3:30 & 1 & Y & N\\
\hline  7 & 2 & 134 & 3:30/24:11 & 2 & Y & N\\
\hline  8 & 2 & 7 & 1:46/2:11 & 2 & Y & Y\\
\hline  9 & 2 & 9 & 6:52/8:32 & 9 & Y & N\\
\hline  10 & 2 & 12 & 5:28/6:31 & 2 & Y & Y\\
\hline  11 & 1 & 2 & 0:24/0:56 & 1 & Y & Y\\
\hline  12 & 1 & 2 & 0:20/0:44 & 1 & Y & Y\\
\hline  13 & 1 & 7 & 0:36/3:40 & 2 & Y & Y\\
\hline  14 & 1 & 7 & 0:30/2:58 & 4 & Y & Y\\
\hline  15 & 1 & 17 & 1:05/8:41 & 2 & Y & Y\\
\hline  16 & 1 & 157 & 0:28/57:19 & 4 & Y & Y\\
\hline
\end{tabular}
\caption{Ocasta recovery performance. For each error, we show the average cluster size, the number of trials required for Ocasta to find the offending cluster using DFS, the recovery time in minutes and seconds to find the offending cluster vs the time for Ocasta to search all the clusters, and the total number of unique screenshots, and the comparison of the effectiveness between Ocasta and Ocasta-NoClust.\label{tbl:effectivenes}}
\vspace{-20pt}
\end{table}

\subsection{Configuration repair}

The traces we collected contain realistic application usage, but because they are collected without interacting with the users of the applications, we are unable to confirm if configuration errors occurred during trace creation.  In addition, we want to be able to precisely control the time at which the configuration error occurs in each trace.  Thus, we simulate configuration errors by injecting a write into the trace at the point in time at which we want the error to occur, that changes the offending setting to the erroneous value. If the configuration error is caused by presence or absence of the offending setting, we insert or delete the setting in the trace.  To simulate the recording phase of Ocasta, we populate the TTKV of the test machine with one of the traces that exhibited usage of the same application in the configuration error scenario.  

We first evaluate how effective Ocasta is at fixing 16 real-world configuration errors, numbered 1-16 in Table~\ref{tbl:misconfig}, which are all configuration errors that were either previously used in the literature~\cite{peerpressure, code} or were found via online forums, FAQ documents and configuration documents. To demonstrate the benefit of using clustering, we compare the effectiveness of Ocasta with the effectiveness of a modified version of Ocasta, called Ocasta-NoClust, that does not use clustering and rolls back a single configuration setting at a time when it tries to fix errors. 

We use as many complex and real configuration errors as possible for the evaluation. For example, error \#12 was found on an internet message board, where the discussion contained 56 messages spanning 3 months. However, we are restricted to only using errors where the offending setting(s) have been modified in our traces -- otherwise Ocasta will have no clustering information for them and Ocasta's repair tool will have no values to roll back to.  This problem cannot happen in practice because any configuration key that is misconfigured must have a modification history on a particular system.  We simulate the configuration error by injecting the erroneous value into the TTKV 14 days before the end of the trace and invoke Ocasta in recovery mode.  For each error, we provide a suitable trial and set the start time to 14 days before the end of the trace.  We configure Ocasta to use the DFS search strategy.  

%Ocasta initially uses a window size of 1 second and a clustering threshold of 2, which will tend to produce smaller, more selective clusters and minimize the number of oversized clusters.  
We evaluated Ocasta using the minimum window size of 1 second and the maximum
correlation threshold of 2, because these produce smaller clusters and are thus the most likely
to lead to invalid configurations or failed fixes. In practice, a user can adjust these settings in case they fail to cluster the configuration settings that cause the configuration problem. With these parameters, Ocasta was able to successfully find the offending cluster and fix the errors in all cases except errors \#2 and \#4.  In both of these cases, the settings that needed to be rolled back were split into several clusters.  In error \#2, the offending settings consisted of one rarely-changing dominant setting, which controls the validity of another 50 settings that change frequently over a moderate span of time, as we described in Figure~\ref{fig:example1}.  When the clustering threshold is reduced to 1, the dominant setting is clustered with 34 of the other settings, but there remain 26 settings that were not clustered together.  When we increase the window size to 30 seconds, causing all settings to be clustered together.  In error \#4, one setting stores an ordered list of names of settings that store applications capable of opening Flash video files.  The setting storing the list tends to change even when the setting storing the application name does not change. Reducing clustering threshold to 1 caused both the setting storing the list and the settings storing application names to be clustered together.

Quantitative results are shown in Table~\ref{tbl:effectivenes}. We can see that Ocasta successfully fixed all 16 configuration errors, but Ocasta-NoClust failed to fix 5 configuration errors, because it requires rolling back more than one configuration settings at a time to fix them. The average cluster size varies between 1 and 8 for our errors, thus effectively reducing the search space by the same factor because Ocasta searches clusters of keys at a time instead of individual keys.  The time column gives the time required by Ocasta to find the offending cluster versus the total time for Ocasta to search all cluster versions up to the 14 day start time.  This shows that Ocasta's sort is successful at prioritizing the clusters, finding the offending cluster by an average of 78\% faster than having to search the entire history.  The screenshots column gives the total number of unique screenshots produced by Ocasta, while the trials column indicates the number of trials executed before the offending cluster is found.  The user must examine an average of 3 screenshots, with a worst case of 11, indicating a very modest amount of user effort.

%We note that in 6 cases where only one unique screenshot is produced that in theory an automated binary search is possible since there is only the erroneous screenshot produced initially by the user when creating the trial and the fixed screenshot.  Unfortunately, this only occurs in 6 of the 16 cases, further motivating the need for Ocasta's unattended search.

\iffalse
\myfig{bydays_all}{Average number of trials as a function of time of the error.}{fig:bydays_all}{0.7}{-10pt}
\myfig{bywrites_all}{Average number of trials as a function of the number of spurious writes.}{fig:bywrites_all}{0.7}{-10pt}
\myfig{bybounds_all}{Average number of trials as a function of the time length defined by the start time.}{fig:start_time}{0.7}{-10pt}
\fi
\mysubfigthree{bydays_all}{By time of errors}{fig:bydays_all}{bywrites_all}{By number of spurious writes}{fig:bywrites_all}{bybounds_all}{By time length}{fig:start_time}{Comparison between DFS and BFS.}{}

%\mysubfigthree{bydays_all}{Time of the error.}{fig:bydays_all}{bywrites_all}{Number of spurious writes.}{fig:bywrites_all}{bybounds_all}{Time bounds.}{fig:start_time}{Average number of trials as the function of a)time of the error, b)number of spurious writes, c)time bounds}{}

Recall that instead of using DFS, Ocasta can also use BFS as the search strategy.  To study the trade-offs we perform searches using both strategies over all 16 errors while varying the number of days in the past when the error was injected, as well as fixing the injection time at 14 days in the past and adding between 0-2 spurious writes after the initially injected error to simulate the case where the user tried to fix the configuration error for 0-2 times.  Figure~\ref{fig:bydays_all} shows the average number of trial executions as a function of error injection time for BFS and DFS.  As can be seen, the number of trials by both BFS and DFS increases as the injection time occurs further in the past, as a result of Ocasta's bias towards checking more recently modified clusters first, while DFS provides better performance overall.  Figure~\ref{fig:bywrites_all} shows the average number of trials as a function of the number of spurious writes after the injected error.  BFS search is highly sensitive to this parameter because to search more writes within a cluster, it must try every other cluster as well, so the number of rollbacks increases if there are a lot of clusters.  

We now evaluate the effect of the start time, which controls the time period Ocasta searches over, on the number of trials Ocasta must execute.  Figure~\ref{fig:start_time} shows the average number of trials Ocasta perform in its search as start time goes further into the past.   As can be seen, the number of trials rises roughly linearly with the length of time the search is conducted over.

\label{sec:quality}
\label{sec:sensitivity}

%A critical requirement of Ocasta is that its clustering should not miss any dependencies that could result in corrupted configuration settings.  We rolled each of our 30 clusters back to a previous snapshot and executed the application.  If the configuration file is corrupted, applications typically either crash, print out an error that the configuration file is corrupted or silently correct the corrupted settings.  In all cases, the application did not exhibit any of these aforementioned behaviors.  From this we surmise that Ocasta's clustering does not miss any dependent keys.

%We manually examine each of the offending clusters to determine whether the keys in it were related or dependent.  {\bf (give counts?)}

\iffalse
\myfig{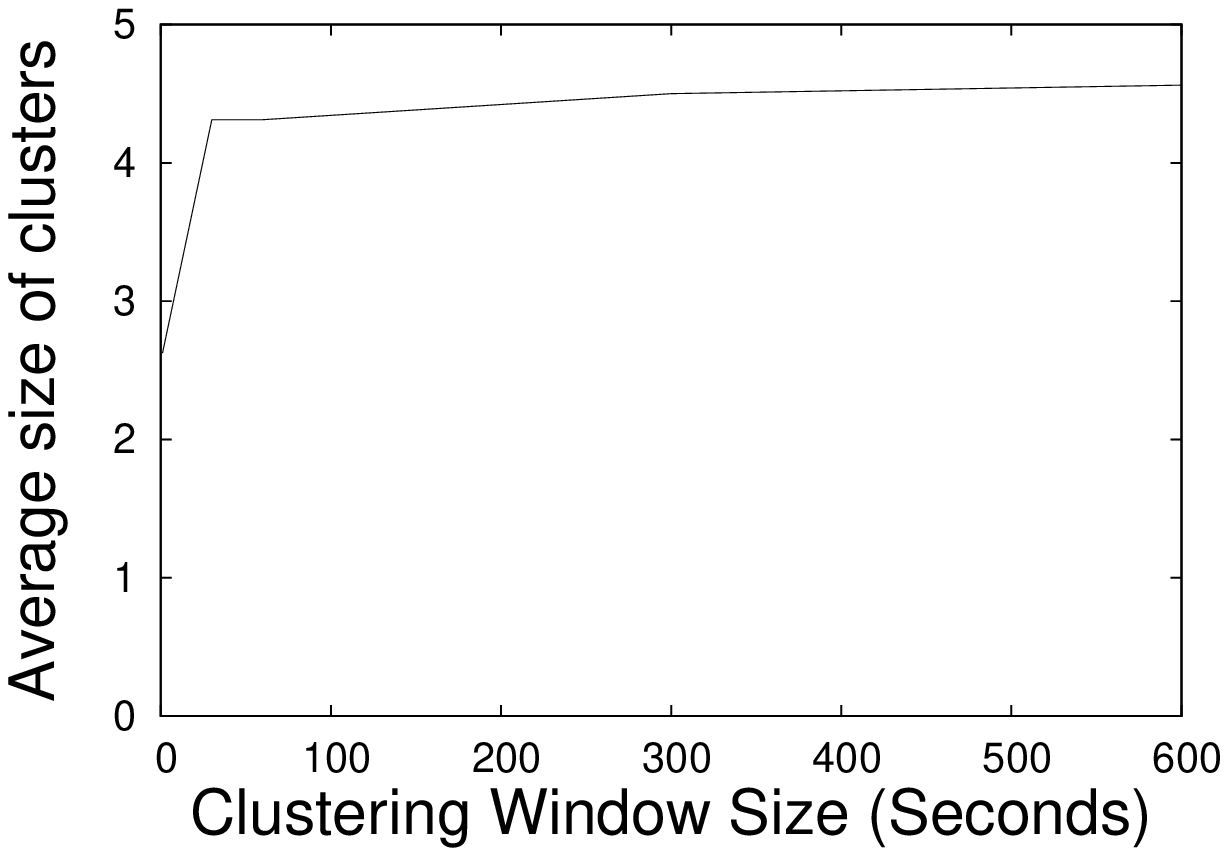}{Average cluster size as a function of window size in our traces.}{fig:window_sensitivity}{0.7}{-10pt}
\myfig{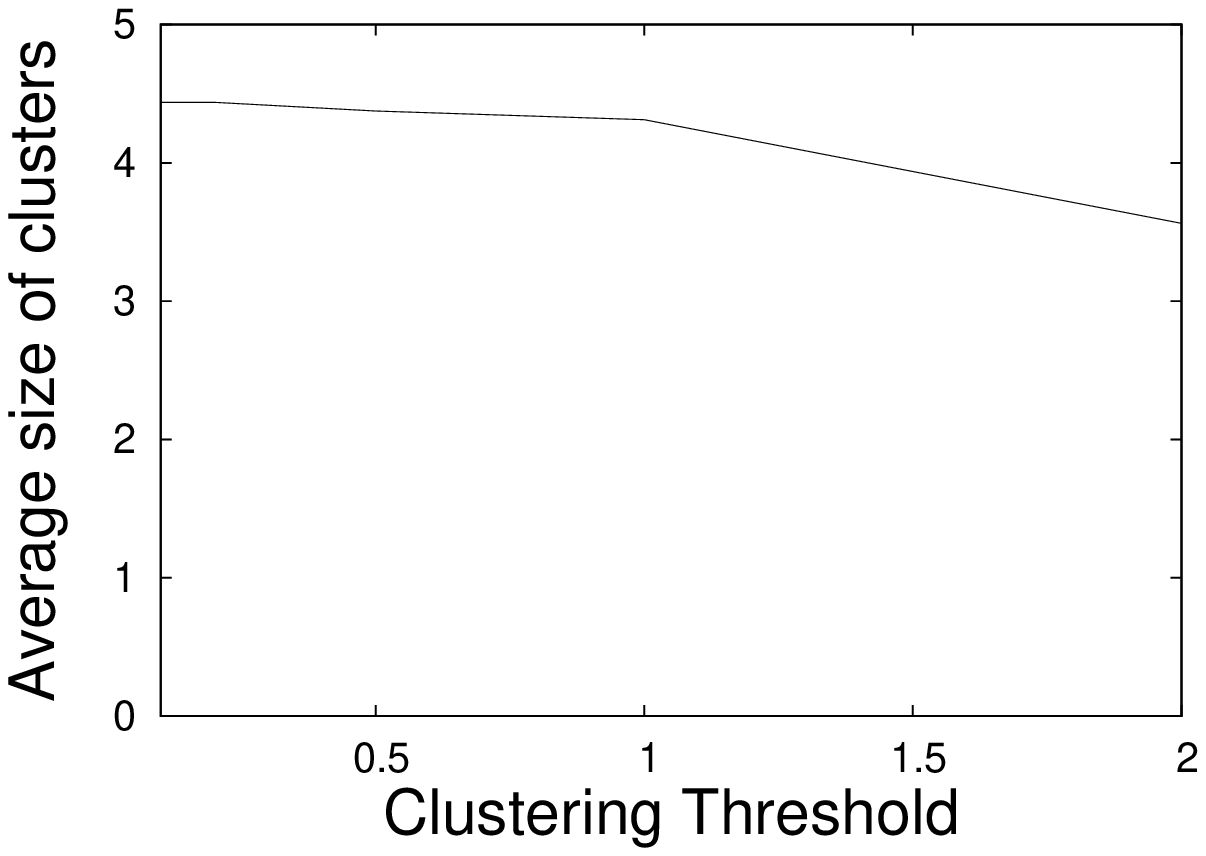}{Average cluster size as a function of clustering threshold in our traces.}{fig:threshold_sensitivity}{0.7}{-10pt}
\else
\mysubfigtwo{windowsize_sensitivity}{Window size.}{fig:window_sensitivity}{threshold_sensitivity}{Clustering threshold.}{fig:threshold_sensitivity}{Average cluster size.}{}
\fi

\subsection{Sensitivity}
We examine the sensitivity of cluster size to both windows size and clustering threshold.  Larger clusters mean fewer trials, but also lead to the potential for more unrelated keys getting changed if the offending cluster grows in size. Figures~\ref{fig:window_sensitivity} and~\ref{fig:threshold_sensitivity} show the growth in average cluster size as a function of both the window size and clustering sensitivity.  The sharp drop at the left hand side of Figure~\ref{fig:window_sensitivity}, is when the window is changed from one second to zero seconds (modifications must have the same timestamp at zero seconds).  Since our traces only record key modification times to the nearest second, there is a lot of noise between these two points.  With the exception of this artifact, the average cluster is relatively insensitive to either parameter, and ranges between between roughly 3.5 to about 4.5 or 25\% of its value.  These results indicate that the overall cluster size is relatively insensitive to changes in these parameters, which might suggest that users should tend to prefer smaller thresholds and larger window sizes to minimize the chances of the offending cluster being undersized.

\subsection{User Study}
To evaluate the effectiveness of the Ocasta repair tool with default settings \footnote{1-second sliding time window, clustering threshold of 2, and DFS search strategy} , we performed a user study on 19 participants with various backgrounds. Because this study contains human subjects, we have obtained a second ethics approval for this study from our institutional ethics review board.  The participants include two faculty members from our department, 13 graduate students from four different departments, a system administrator, an administrative assistant, and two software engineers.  Six out of the 19 participants of the user study are non-technical
users. None of participants were authors of this paper and none were compensated for this user study.  Each participant was given a brief explanation on how Ocasta works and shown a demonstration on a contrived configuration error.  The participant then tested Ocasta on a computer setup with configuration error \#11, \#13, \#15 and \#16 from Table \ref{tbl:misconfig}. We use only four errors to limit the length of the user study, because it took between 1.5 and 2 hours for each participant to finish the user study. In each case, the participants were first asked to quantitatively rate how familiar were they with the application having the configuration error. Then they were given a description of the error and were asked to use Ocasta to fix the configuration error. We recorded the time the participants took to create the trial.  After they finished creating the trial, they were asked to quantitatively rate how difficult it was to produce the trial.

The participant was then shown the set of screenshots Ocasta produces when run on the history from our traces and asked to select the screenshot that showed the fixed application.  The time taken for the participant to select the screenshot was also recorded.  After the participant selected the screenshot, we recorded whether they selected the right one.  We also asked the participant how many of the screenshots they actually examined and to qualitatively rate how difficult it was to find the screenshot.   

We then reset the system back to its misconfigured state and asked the participant to try to fix the error manually.  The participant was given full control of the computer and was allowed to use Internet to search for possible solutions to the configuration error. To keep the test short, we cut the participants off at 5 minutes.  We recorded whether the participant was able to fix the error manually or not and the time it took for them to fix the error.  For each error, the participant was finally asked whether they had experienced the particular error themselves before and the steps they took to fix or try to fix the error. 

\myfig{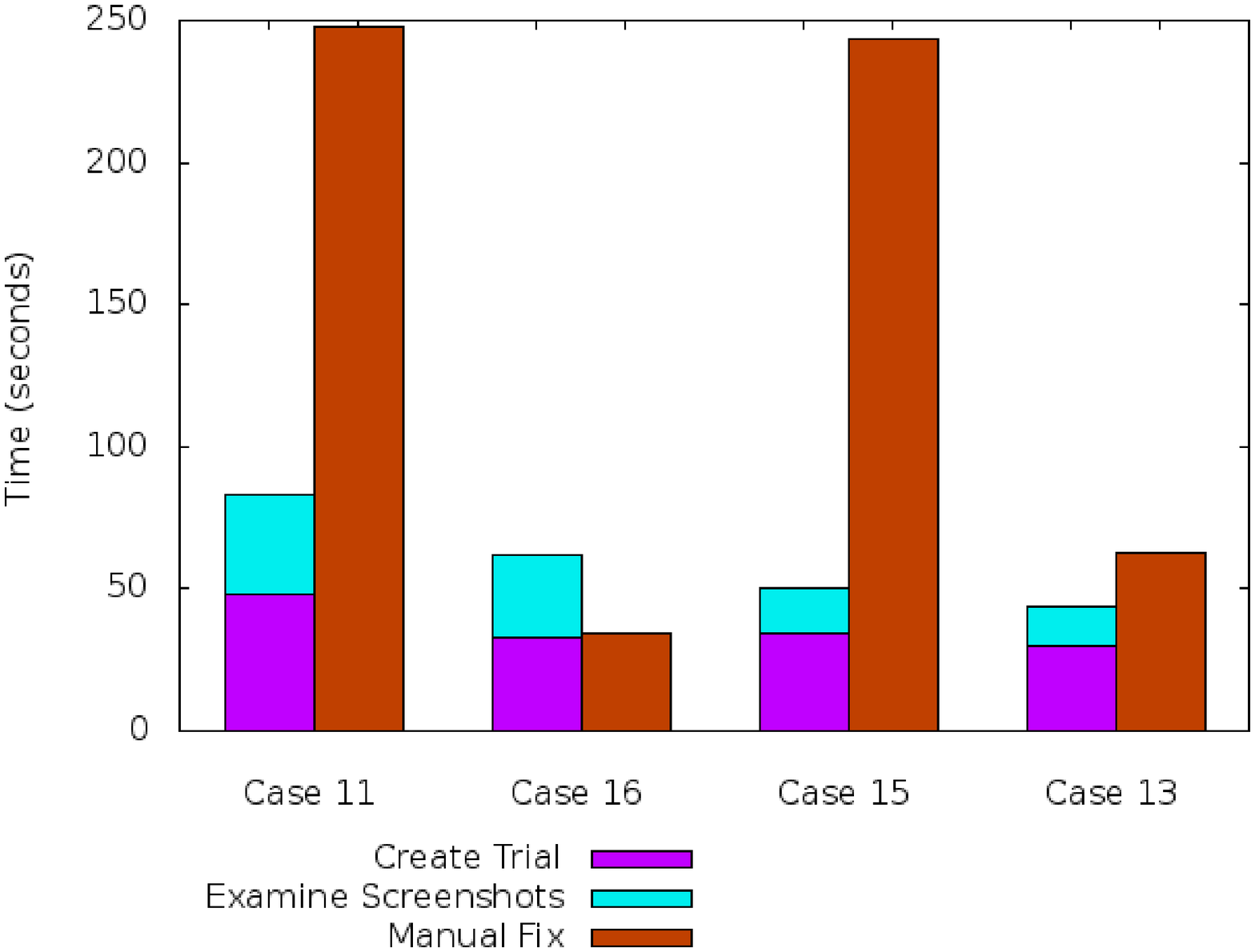}{Comparison of time required to fix the error with Ocasta versus manual fixing from our user study.}{fig:user_study}{0.7}{-10pt}

Figure~\ref{fig:user_study} shows a comparison between the average time users took to both create the witness and select the screenshot and the average time taken to manually repair each configuration error.  If we use the time spent as an indicator of the amount of user effort, we can see that Ocasta saves users a significant amount of effort to repair configuration errors.  Only in case 16 were the majority of participants able to fix the configuration error manually and this significantly lowered the average time for the a manual fix.  Qualitatively on a difficulty scale of 1 to 5, with 1 being the easiest, across the 4 errors, the participants rated the creation of the trial as 1 74\% of the time, 2 21\% of the time and and 3 5\% of the time.  For selecting the correct screenshot, participants rated the difficulty as 1 80\% of the time, 2 11\% of the time, 3 8\% of the time and 4 1\% of the time.  

Our user study has several sources of bias.  First, selection of participants was not completely random, but consisted of colleagues and acquaintances of the authors.  Second, the administration of the study was single blind and the person administrating the test knew the correct answer.  To minimize this effect, we tried to minimize interaction with the participant and communicated using written materials as much as possible.  Third, the participants were cut off at 5 minutes when they tried to fix the error manually, while no cut off was used for generating the Ocasta trial or selecting the screenshot.  Thus, the time measurements for some of the manual fixes represent a lower-bound while the time measurements for Ocasta usage are precise.  Finally, we selected errors that tended to be simple.  This made it easier to explain the errors to users who might be unfamiliar with the applications.  In addition, simple errors make manual fixing easier and thus make it more difficult for Ocasta to have a significant advantage over manually searching for the fix.

\section{Related Work}\label{sec:related}

%\subsection{Troubleshooting Configuration Errors}
%There are essentially three broad areas of work related to Ocasta.  The most closely related work deals with repairing or troubleshooting configuration errors.  Ocasta is also related to systems that attempt to detect configuration errors.  Finally, there is also work that focuses on using roll back for system recovery and debugging.

\paragraph{Inferring related configuration settings} Few previous studies automatically infer relations among configuration settings. Zheng et al. \cite{Zheng2007} deduce dependency among configuration settings by experimentally testing the impact of changing configuration settings. Ocasta's clustering algorithm avoids the overhead of experimental tests by using observed application accesses to configuration settings. Glean~\cite{glean} infers relations among configuration settings by analyzing hierarchical structure of configuration settings, while Ocasta's clustering algorithm does not require the existence of hierarchial structure for configuration settings. 

\paragraph{Diagnosing configuration errors} Of the work that focuses on diagnosing configuration errors, Ocasta is most closely related to Strider~\cite{strider} and PeerPressure~\cite{peerpressure}. Both PeerPressure and Strider use a genebank of common configurations and apply statistical methods to determine where the error might lie.  These systems assume homogeneity across machines and also have privacy implications as users must share their configurations with the genebank.  Ocasta only requires information collected locally from the machine with the error and thus does not have the drawbacks of a genebank.  

ConfAid~\cite{confaid} takes a ``white-box'' approach by using taint-analysis to try to identify the configuration setting that causes a configuration error.  ConfAid ranks configuration settings that affect the path taken to reach the configuration error as more likely to be configuration keys that can fix the error. Another ``white-box'' approach, Failure-Context-Sensitive analysis \cite{Rabkin2011} extracts the mapping between configuration settings and the source code lines that can be affected by these configuration settings, from the source code of an application. These mappings can be used to identify the configuration setting that causes configuration errors, when the source code lines of the errors are available, for example from an application's error message. More recent work, ConfDiagnoser, combines static analysis of an application's source code and execution profiling to rank configuration settings that causes executions to deviate from pre-generated correct executions \cite{Zhang2013}.  Because these approaches are white-box, they require application source code.  In contrast, Ocasta treats applications as black-boxes and only requires access to the application's key-value store. 

All above work focuses on identifying a single configuration setting that causes configuration errors. With the clustering provided by Ocasta, their techniques can be leveraged to diagnose configuration errors caused by more than one configuration settings.

Chronus~\cite{chronus} maintains a history of entire system states and focuses on using binary search to find the optimal recovery point in an application's history. Chronus logs at the disk block layer and as a result, many of the historical states it generates can corrupt file systems and thus cannot be used for recovery.

\paragraph{Fixing configuration errors} Kardo~\cite{kardo} and Autobash~\cite{autobash} are both systems that take a human-generated solution for a configuration error, perform analysis on the solution to find the minimum set of actions that make up the configuration fix and generalize it so it can be applied to a wider set of machines. Ocasta does not require human-generate solutions.

\paragraph{Detecting configuration errors}  Like Ocasta, CODE~\cite{code} analyzes the accesses patterns that applications make to the Windows registry.  CODE uses a rule learning algorithm to identify normal key access patterns of an application and flags anomalous access patterns as possible configuration errors.  CODE detects configuration errors, but unlike Ocasta, it does not fix the errors, nor does it try to identify relationships between keys other than the access patterns.  Conferr is a tool for quantifying system manageability and resilience to configuration errors~\cite{conferr:dsn, conferr:hotdep}.  It uses simulated human models to try to generate realistic configuration errors. Both CODE and Conferr can be viewed as complementary to Ocasta.

\paragraph{Time travel and roll back} The concept of time travel and roll back has been used for debugging and system recovery from intrusions.  Time-travel virtual machines~\cite{TTVM} enables deterministic replay of whole machines to simplify OS debugging. Taser~\cite{taser} and Retro~\cite{retro} use system-level tracking and perform selective recovery after an intrusion.  Rx~\cite{Feng2005} uses repeated roll backs to find an execution where bugs do not occur, but does not try to find the root cause or attempt to permanently fix the bug.  Like Ocasta, these systems use roll back recovery but focus on fixing other types of faults while Ocasta focuses specifically on configuration errors.

\paragraph{Hierarchical clustering} Many previous studies have used hierarchical clustering for software clustering~\cite{Schwanke1991, Deursen1999, Andritsos2005}, including program comprehension, reverse engineering, and software reengineering, cluster different levels of abstractions of software artifacts, such as variables, functions, and source files. Prior work has also used hierarchical agglomerative clustering to improve the efficiency of finding software failures during software testing~\cite{Dickinson2001} or categorizing software failures~\cite{DiGiuseppe2012}. They cluster profiles of an application's executions. 

Ocasta uses the maximum linkage criterion, which as been found by other prior work~\cite{Anquetil1999, Maqbool2007} to provide better performance than other linkage criterion. Ocasta augments the hierarchical agglomerative algorithm to be able to partition clusters using an adjustable clustering threshold, which is more flexible and intuitive for our purposes of clustering configuration settings.  

\section{Conclusion}\label{sec:conclusion}

We describe the design and implementation of Ocasta, a system that enables configuration recovery systems to handle multi-configuration setting errors by identifying clusters of related configuration settings using statistical clustering.   We have evaluated Ocasta over several months on both Windows and Linux machines and find that Ocasta's clustering accurately identifies about 88.6\% of clusters on average. Our evaluation of Ocasta in fixing configuration errors shows that Ocasta successfully fixed all 16 real world configuration errors used in our evaluation, 5 of which require changing more than one configuration setting together to fix, by utilizing the identified clusters of related configuration settings, 

\section*{Acknowledgments}

We thank Ding Yuan for his invaluable suggestions on our user study and Tim Trant for helping us setting up our trace collection infrastructure. We also thank the anonymous reviewers for their helpful comments. This research was partially supported by an ORF-RE grant from the Ontario Ministry of Research and Innovation and by an NSERC Discovery Grant.

%Remember to thank those people who provided resources, feedback or guidance, 
%especially those who funded the work. 

%The asterisk after the section specification makes this section unnumbered.

%\section*{Availability}

%It's great when this section says that MyWonderfulApp is free software, 
%available via anonymous FTP from
%\begin{center}
%{\tt ftp.site.dom/pub/myname/Wonderful}\\
%\end{center}

%Also, it's even greater when you can write that information is also 
%available on the web at

%\begin{center}
%{\tt http://www.site.dom/\~{}myname/SWIG}
%\end{center}

%Now we get serious and fill in those references.  Remember you will
%have to run latex twice on the document in order to resolve those
%cite tags you met earlier.  This is where they get resolved.
%We've preserved some real ones in addition to the template-speak.
%After the bibliography you are DONE.

%{
%
%  \bibliographystyle{acm}
%  \bibliography{bibfile}
%}
\bibliographystyle{IEEEtran}
\bibliography{IEEEabrv,bibfile}
\end{document}